\documentclass[11pt, a4paper]{article}
\pdfoutput=1

\usepackage{jcappub}
\usepackage{subfigure}
\usepackage{bm}
\usepackage{amssymb}
\usepackage{epsfig}
\usepackage{float}
\usepackage{caption}
\usepackage{color}
\usepackage{upgreek}
\usepackage{amsmath}
\usepackage{graphicx}
\usepackage{float}
\usepackage[utf8]{inputenc} 

\newcommand{\be}{\begin{equation}}
\newcommand{\ee}{\end{equation}}
\newcommand{\bea}{\begin{eqnarray}}
\newcommand{\eea}{\end{eqnarray}}

\newcommand{\Rm}[1]{\mathrm{#1}}
\newcommand{\Bf}[1]{\bm{#1}}

\begin{document}

\title{Multi-Scale Perturbation Theory II: \\Solutions and Leading-Order Bispectrum in the $\Lambda$CDM Universe}

\author[a]{Christopher S. Gallagher,}
\emailAdd{c.s.gallagher@qmul.ac.uk}

\author[a]{Timothy Clifton,}
\emailAdd{t.clifton@qmul.ac.uk}

\author[a,b,c]{Chris Clarkson}
\emailAdd{chris.clarkson@qmul.ac.uk}

\affiliation[a]{School of Physics and Astronomy, Queen Mary University of London, UK.}
\affiliation[b]{Department of Physics \& Astronomy, University of the Western Cape, Cape Town 7535, South Africa}
\affiliation[c]{Department of Mathematics \& Applied Mathematics, University of Cape Town, Cape Town 7701, South Africa}


\abstract{
Two-parameter perturbation theory (2PPT) is a framework designed to include the relativistic gravitational effects of small-scale nonlinear structures on the large-scale properties of the Universe. In this paper we use the 2PPT framework to calculate and study the bispectrum of matter in a spatially-flat $\Lambda$CDM cosmology. This is achieved by deploying Newtonian perturbation theory to model the gravitational fields of quasi-nonlinear structures, and then subsequently using them as source terms for the large-scale cosmological perturbations. We find that our approach reproduces some of the expected relativistic effects from second-order cosmological perturbation theory, but not all. This work therefore provides a first step in deploying a formalism that can simultaneously model the weak gravitational fields of both linear and nonlinear structures in a realistic model of the Universe.
}

\maketitle
\flushbottom

\bibliographystyle{unsrt}


\section{Introduction}

The Universe we inhabit contains nonlinear structures on small scales ($\lesssim 100 \, {\rm Mpc}$) \cite{blanton2017sloan}, and linear structures on large scales ($\gtrsim 100 \, {\rm Mpc}$) \cite{aghanim2020planck}. The gravitational physics in these two different regimes can both be understood by using weak-field expansions of Einstein's field equations, assuming one does not wish to resolve compact objects like neutron stars and black holes \cite{poisson2014gravity}. However, the precise nature of the approach that is required to accurately model each regime is not the same.

On small scales cosmologists routinely appeal to the post-Newtonian limit of Einstein's theory, as Newton's law of gravity is capable of modelling the gravitational fields produced by arbitrarily large density constrasts (assuming gravity remains weak, and the speed of matter fields remain non-relativistic) \cite{will2018theory}. However, on large-scales the Newtonian approach cannot be applied, and cosmological perturbation theory must be used to model the gravitational fields of matter \cite{malik2009cosmological}. Cosmological perturbation theory is extremely versatile, but is only applicable in the limit where density constrasts and the peculiar velocity of matter fields are both small (of the order of magnitude as the gravitational potentials themselves, i.e. $\sim 10^{-5}$).

The usual approach to dealing with the different mathematical expansions required in each of these two limits is to assume that each can be performed independently of the other, and that the results of both should overlap in some intermediary regime. While this is plausible at linear order, it is unlikely to be a viable approach at second and higher order in either of the two expansions. This is because second- (and higher-) order field equations are known to mix scales. That is, functions that have support only on small scales only can source large-scale fluctuations when combined quadratically, or in higher powers, with other perturbations.

The ``2-Parameter Perturbation Theory'' (2PPT) was recently introduced in order to provide a mathematical framework in which to understand and study this problem \cite{goldberg2017cosmology, goldberg2017perturbation}. This approach simultaneously performs a (post-)Newtonian expansion and an expansion in cosmological perutrbation theory around the same Friedmann background. It makes explicit which features of each expansion can be understood as being independent of the other, as well as the ways in which the two expansions are linked and interact. In particular, this approach makes explicit how the Friedmann equations of the background cosmology are sourced by the average of the Newtonian masses that exist within it, and how the linear perturbation equations for large-scale perturbations can be sourced by quadratic and cubic combinations of the Newtonian quantities from the small-scale sector of the expansion \cite{carlson2009critical}.

In a previous paper we examined the 2PPT approach in the test case of a universe with vanishing cosmological constant (i.e. with $\Lambda=0$), and during the epochs in which the gravitational effects of relativistic matter are negligible \cite{gallagher2020multi}. The leading-order equations for describing the small-scale physics in this case are identical to the usual equations used to describe Newtonian gravity on an expanding background \cite{bagla2005cosmological}, and were solved using Newtonian perturbation theory \cite{bernardeau2002large}. These solutions were then used to solve the equations for the leading-order part of the gravitational fields on large scales, order-by-order in the Newtonian perturbation theory expansion.

It was found that when $\Lambda=0$ the leading-order contribution to the 2PPT equations precisely reproduces the expected equations from cosmological perturbation theory and Newtonian gravity, within their respective domains of applicability. The next-to-leading order contributions to the 2PPT equations, which describe the evolution of long wavelength linear perturbations on top of an FLRW background that has been allowed to develop short-scale Newtonian nonlinear inhomogeneities, were then investigated. These contributions provide terms that appear in the calculation of the bispectrum of matter fluctuations, and were shown to produce sizeable corrections to the Newtonian bispectrum on scales $k \lesssim 10^{-2} \, {\rm Mpc}$ \cite{gallagher2020multi}. They include most (though not all) of the terms that occur in second-order cosmological perurbation theory, and give the leading-order relativistic effect of small-scale nonlinearities on the matter bispectrum.

In this paper we generalize the study performed in Ref. \cite{gallagher2020multi} to Friedmann cosmologies with non-zero cosmological constant, $\Lambda \neq 0$. Such situations are of obvious significance for observational cosmology, and require separate study due to the extra complexity that a non-zero $\Lambda$ introduces into the evolution of perturbations. The mathematical foundations of including $\Lambda\neq 0$ in the 2PPT approach have already been investigated in Ref. \cite{goldberg2017perturbation}, where the gauge depenencies and field equations that result were thoroughly investigated. Here we will use these results to calculate the leading-order relativistic corrections to the matter bispectrum within the 2PPT formalism, and with a non-vanishing cosmological constant.

We find that the time-dependence that the presence of a non-zero cosmological constant induces into the first-order gravitational potentials significantly complicates the application of the 2PPT equations, and increases the disparity between this approach and the more traditional use of second-order cosmological perturbation theory. While the 2PPT equations faithfully include the effects of small-scale nonlinearities on the relativisitic bispectrum, they do not include the second-order effects of large-scale linear fluctuations, which are non-negligible on very large scales. This demonstrates that the 2PPT approach may need to be supplemented by additional terms in order to reliably include all relativistic effects in the calculation of at least some cosmological observables on very large scales.

In Section \ref{2pptsec} we explain the 2PPT formalism, as required for use in a $\Lambda$CDM universe. In Sections \ref{sec:sols} and \ref{sec:kernels} we then solve these equations using Newtonian perturbation theory. This is followed by a calculatation of the bispectra of gravitational potentials and dark matter in Section \ref{bispectrumsec}, and a conclusion in Section \ref{sec:dis}. We use Latin letters for spatial indices and dashes to refer to differentiation with respect to conformal time. Spatial derivatives are written $\partial_i$, and $\nabla^2$ refers to the Laplacian operator associated with these derivatives. We also choose to work in geometrized units $G = c = 1$, and in longitudinal gauge \cite{clifton2020viable}, throughout.

\section{Two-parameter perturbation theory in $\Lambda$CDM}
\label{2pptsec}

Two-parameter perturbation theory is constructed by performing post-Newtonian and cosmological perturbation theory expansions around a single Friedmann background, which results in a line-element that can be written in longitudinal gauge as \cite{goldberg2017perturbation}
\begin{equation} \label{ds2}
ds^2 = a(\tau) \bigg[-(1+2U + 2\phi) d\tau^2 + (1-2U - 2\psi) \delta_{ij} dx^i dx^j \bigg] \, .
\end{equation}
In what follows, we will briefly outline the equations that must be obeyed in the presence of a non-zero $\Lambda$ by the Newtonian potential $U$, the scale factor $a$, and the large-scale cosmological perturbations $\phi$ and $\psi$. For further details, and more thorough mathematical treatment, justification and introduction, the reader is referred to Ref. \cite{goldberg2017perturbation}.

\subsection{Small-scale Newtonian equations}
\label{newtss}

The $U$ in this expression corresponds to the Newtonian gravitational potential, which is the leading-order contribution to the line-element from the post-Newtonian expansion, and which satisfies the usual Newton-Poisson equation on an expanding background:
\begin{equation} \label{newtU}
\nabla^2 U = 4\pi a ^2 \bar{\rho}\, \updelta_{\Rm{N}} \, ,
\end{equation}
where $\bar{\rho} = \langle \rho_{\rm N} \rangle$ is the cosmological average of the Newtonian mass density (taken over a suitable domain), and $\updelta_{\rm N}=\rho_{\rm N}/\bar{\rho}_{\rm N}$ is the density contrast in this quantity. The evolution equations for $\updelta_{\rm N}$ is given by the energy conservation equation,
\begin{align} 
\updelta_{\Rm{N}}' + \theta_{\Rm{N}} &=- \partial^i \big(  \updelta_{\Rm{N}} v_{{\Rm{N}}i}\big) \;, \label{lcdmcont}
\end{align}
where $v_{{\Rm{N}}i}$ is the 3-velocity of the Newtonian masses (assumed to be of order $v^2 \sim U$), which has its own evolution equation as follows:
\begin{align} 
\theta_{\Rm{N}}' + \mathcal{H} \theta_{\Rm{N}}+ \frac{3\mathcal{H}^2}{2}\updelta_{\Rm{N}} &= - \partial^i\big( v_{{\Rm{N}}j}\partial^j v_{{\Rm{N}}i} \big) \;, \label{lcdmeuler}
\end{align}
where $\theta_{\Rm{N}} = \partial^i v_{\Rm{N}i}$ is the corresponding velocity divergence. The relative size of the terms $\updelta_{\Rm{N}}'$, $\theta_{\Rm{N}}'$ and $\mathcal{H}$ are determined in this expansion by noting that time derivatives add an order of smallness $\sim v$ compared to spatial derivatives (see Ref. \cite{goldberg2017cosmology}).

These equations are all identical to the usual expressions for Newtonian gravity and dynamics on an expanding background \cite{bagla2005cosmological}, but here are formally derived as the leading-order parts of a relativistic post-Newtonian expansion on an expanding background \cite{milillo2015missing}, which is used to describe gravitational fields on small scales. The reader will note that there is no restriction implied on the magnitude of the quantity $\updelta_{\rm N}$ in these equations, though the average of the density is found to be of order $\bar{\rho}\sim U$. This means that $\updelta_{\rm N}$ can be used to describe highly nonlinear structures. It may also be noted that the cosmological constant, $\Lambda$, does not appear in these equations directly, and only influences $U$ through the effect it has on the scale factor, $a(\tau)$.

\subsection{Dynamical equations for the background}
\label{backgroundss}

The background dynamics in the 2PPT approach are given by equations that are obtained by averaging the Newtonian-level contributions to the field equations, which results in the effective Friedmann equations \cite{goldberg2017perturbation}
\begin{align} \label{leadingordersep}
\mathcal{H}^2 &= \frac{8\pi a^2}{3} \bar{\rho}+ \frac{\Lambda}{3} a^2 \;, \\[5pt]
\mathcal{H}' &= -\frac{4\pi a^2}{3} \bar{\rho}  + \frac{\Lambda}{3} a^2 \, ,
\end{align}
where $\mathcal{H}=a'/a$, and where primes denote differentiation with respect to conformal time $\tau$. The quantity $\bar{\rho}$ that appears in these equations is the average of the Newtonian mass density, which obeys the conservation equation \cite{goldberg2017cosmology}
\begin{equation}
\label{encon}
\bar{\rho}'+3 \mathcal{H} \bar{\rho} =0 \, ,
\end{equation}
where time derivatives have again been used to assign an extra order of smallness of order $\sim v$, and where we have taken $\Lambda \sim v^2$.

The reader may note that the density $\bar{\rho}$ that appears in these equations is the direct result of integrating the nonlinear Newtonian mass density $\rho_{\rm N}$ over large cosmological domains, and is not a separate component of the energy density (as it is in cosmological perturbation theory). Otherwise, these equations are identical to those of Friedmann cosmology containing dust and a cosmological constant, and correspondingly have the same solutions.

\subsection{Large-scale cosmological perturbations}
\label{cptss}

The gravitational potentials $\phi$ and $\psi$, which appear in (\ref{ds2}), correspond to the gravitational fields generated by the low amplitude density contrasts $\updelta \ll 1$ that appear on scales $k^{-1} \gtrsim 10^2 \, {\rm Mpc}$. Given the existence of the small-scale Newtonian perturbations on a Friedmann background, and assuming $U \sim \phi \sim \psi$, gives the following evolution equation for these fields \cite{goldberg2017perturbation}:
\begin{align} 
(\psi + U)'' + 3\mathcal{H}(\psi + U)' + a^2 \Lambda (\psi + U) 
=& \frac{4\pi a^2 \bar{\rho}}{3}(1+\updelta_{\Rm{N}})v_{\Rm{N}}^2 + \mathcal{H}(\psi' - \phi') + \frac{1}{3}\nabla^2(\psi - \phi) \nonumber \\&+ \frac{7}{6}(\nabla U)^2 + \frac{2}{3}(\phi + \psi + 2U)\nabla^2U + a^2 \Lambda(\psi - \phi)\;, \label{evollcdm}
\end{align}
and the constraints
\begin{align}
&\partial^i \partial_j (\psi - \phi) + 2 \partial^i U \partial_j U + 2 (\psi + \phi + 2 U) \partial^i \partial_j U 
\nonumber\\
&\hspace{1cm}- \frac{1}{3} \delta^i_{\;j} \bigg[ \nabla^2  (\psi - \phi) + 2  ( \nabla U )^2  + 2 (\psi + \phi + 2 U) \nabla^2 U \bigg] 
\nonumber \\
&\hspace{3cm}= 8\pi a^2\bar{\rho} \; (1 +  \updelta_{\Rm{N}})\big( v_{\Rm{N}}^i v_{\Rm{N}j} - \frac{1}{3}  \delta^i_{\;j}  v_{\Rm{N}}^2 \big) \label{tracefreeij} \, ,
\end{align}
\begin{align}
&\frac{1}{3} \nabla^2 \psi - \mathcal{H}(\psi' + U') - \mathcal{H}^2(\phi+U) 
\nonumber \\
&\hspace{2cm}= \frac{4\pi a^2 \bar{\rho}}{3}\updelta + \frac{4\pi a^2 \bar{\rho}}{3}(1 + \updelta_{\Rm{N}})v_{\Rm{N}}^2  - \frac{1}{2} (\nabla U)^2 - \frac{4}{3}(\psi + U) \nabla^2 U \; ,  \label{genPoisson}
\end{align}
and
\begin{align}
\partial_i \big( \psi'  + \mathcal{H}\phi \big)  = - \frac{3\mathcal{H}^2}{2} (1 + \updelta_{\Rm{N}}) v_{i} \; \label{momentum} \;,
\end{align}
where $v_i$ is the 3-velocity associated with the large-scale perturbations $\updelta$ (not to be confused with the nonlinear small-scale density contrast $\updelta_{\rm N}$).

These equations have been derived under the assumption that vector and tensor modes can be neglected, and exist at a higher order in the 2PPT perturbative expansion than the previous sets presented in Sections \ref{newtss}-\ref{backgroundss}. They combine both large and small scale gravitational potentials, though the small-scale potentials $U$ are already specified by the Newton-Poisson constraint equation (\ref{newtU}). They can also be seen to reduce to the linearized scalar equations from cosmological perturbation theory \cite{malik2009cosmological}, in the case where $U$ vanishes and quadratic products of density constrasts and 3-velocities are neglected.

In the absense of their neglect, there can be seen to be terms that are quadratic and cubic in quantities that would normally be considered to be small in cosmological perturbation theory. These terms appear in these equations because they are derived from the post-Newtonian sector of the theory, and because they are required to not be small in order to recover the appropriate Newtonian limit. Consistent application of the exact same logic forces these terms to appear in Equations (\ref{evollcdm})-(\ref{momentum}), where they can act as sources for the large-scale gravitational potentials $\phi$ and $\psi$. 

While equations (\ref{evollcdm})-(\ref{momentum}) are intended to give the leading-order part of the field equations for the large-scale gravitational potentials in 2PPT, it is instructive to be able to compare them to the equations that one obtains from performing an expansion of the field equations up to second order in cosmological perturbation theory. In this case, it can readily be seen that there are both extra terms in the equations above, when compared to the second-order cosmological perturbation theory equations, as well as missing terms. In particular, equation (\ref{evollcdm}) does {\it not} contain the terms $\frac{1}{2} (U')^2$, $4 \mathcal{H} U \, U'$ and $2 a^2 \Lambda U^2$, and equation (\ref{genPoisson}) does {\it not} contain the terms $-2 \mathcal{H}^2 U^2$ and $-\frac{1}{2} (U')^2$. This is because, in every case, these terms appear at order $\sim v^6$ in 2PPT expansion, which is smaller than every other term in these equations (which are all $\sim v^4$). This difference originates from the fact that time-derivatives add an order of smallness in this approach, as is required for self-consistency of the post-Newtonian limit of this theory (see Ref. \cite{goldberg2017cosmology} for details).

On the other hand, equations (\ref{evollcdm})-(\ref{momentum}) contain terms that would not only normally appear at second-order in cosmological perturbation theory, but also a handfull of terms that would not appear until third order. These include the term $\displaystyle {4 \pi} a^2 \bar{\rho}\, \delta_{\rm N}\, v_{\rm N}^2/3$ in equation (\ref{evollcdm}), the term $8 \pi a^2 \bar{\rho} \, \delta_{\rm N} (v_{\rm N}^i v_{{\rm N} j} - \frac{1}{3} \delta^i_{\phantom{i} j} v_{\rm N}^2)$ in equation (\ref{tracefreeij}), and $\displaystyle {4 \pi} a^2 \bar{\rho}\, \delta_{\rm N}\, v_{\rm N}^2/3$ in equation (\ref{genPoisson}). The appearance of these terms at leading-order in the 2PPT approach demonstrates the fact that $\delta_{\rm N}$ is allowed to be non-perturbatively large in this approach, as is required for consistent inclusion of non-linearities in the density field. We refer the reader to Refs. \cite{goldberg2017cosmology, goldberg2017perturbation} for a more detailed explanation of the derivation of these equations, as well as for mathematical and physical justification of the assumptions that go into them. 

Here we will proceed to find solutions to equations (\ref{evollcdm})-(\ref{momentum}) by using Newtonian perturbation theory to appropriately expand them, as we will now describe. Once we have found solutions, we will compare these to the corresponding equations from second-order cosmological perturbation theory, which we have verified are obtained precisely by re-introducing the terms discussed above into equations (\ref{evollcdm})-(\ref{momentum}), and which can be found in Ref. \cite{bartolo2006full}.

\section{Solutions of the two-parameter equations}
\label{sec:sols}

The order in which the 2PPT equations from Section \ref{2pptsec} need to be solved goes as follows:
\begin{itemize}
\item[(i)] Solve the effective Friedmann equations for the background evolution to find $a(\tau)$; i.e. solve equations (\ref{leadingordersep})-(\ref{encon}) from Section \ref{backgroundss}.
\item[(ii)] Use the scale factor $a(\tau)$ from step (i) to solve the equations for Newtonian gravity on an expanding background; equations (\ref{newtU})-(\ref{lcdmeuler}) from Section \ref{newtss}.
\item[(iii)] Use the solutions from steps (i) and (ii) to solve for the large-scale gravitational potentials  $\phi$ and $\psi$; using equations (\ref{evollcdm})-(\ref{momentum}) from Section \ref{cptss}.
\end{itemize}

The first step here is straightforward, with known exact solutions available for spatially flat FRW cosmologies with dust and $\Lambda$ \cite{aldrovandi2006analytic}. The second step is more complicated, and could be approached in a number of ways \cite{bagla2005cosmological, bernardeau2002large}.  In this paper we use Newtonian perturbation theory to model the solutions to these equations, in order to get approximate solutions for step (ii). This will largely follow the treatments given in Refs. \cite{goroff1986coupling, bernardeau2002large, gallagher2020multi}. With solutions for the $U$, $\updelta_{\rm N}$ and $v_{{\rm N}i}$ in hand, we will then complete step (iii) by solving equations (\ref{evollcdm})-(\ref{momentum}) order-by-order in the Newtonian perturbation theory expansion. This will involve borrowing techniques from second-order cosmological perturbation theory, as explained in Ref. \cite{gallagher2020multi}.

Each step in the method described above will require the specification of appropriate initial conditions, which we will discuss as we proceed. Once we have solutions on both large and small scales, we will proceed to calculate the matter bispectrum of perturbations in Section \ref{bispectrumsec}. This is the simplest statistic that can be constructed that will have non-trivial contributions from the extra term in equations (\ref{evollcdm})-(\ref{momentum}). For the rest of this section, we will proceed to find solutions at first and second order in the Newtonian perturbation theory expansion, which we will refer to as the `first' and `second' approximations.

\subsection{First approximation}

The Newtonian perturbation theory expansion we wish to use decomposes $\delta_{\rm N}$, $\theta_{\rm N}$ and $U$ into linear, second-order, and higher-order parts as
\begin{align} \label{Npertserieslcdm}
\updelta_{\Rm{N}} &= \updelta_{\Rm{N}}^{(1)} + \frac{1}{2}\updelta_{\Rm{N}}^{(2)} + \dots = \sum_{n=1}^{\infty} \frac{\updelta_{\Rm{N}}^{(n)}}{n!} \,,\\
\theta_{\Rm{N}} &= \theta_{\Rm{N}}^{(1)} + \frac{1}{2} \theta_{\Rm{N}}^{(2)} + \dots = \sum_{n=1}^{\infty} \frac{\theta_{\Rm{N}}^{(n)}}{n!} \,,\\
U &= U^{(1)} + \frac{1}{2} U^{(2)} + \dots = \sum_{n=1}^{\infty} \frac{U^{(n)}}{n!}  \;,
\end{align}
where the label in brackets is denoting the order of a quantity in this expansion. These expressions are then subsituted into Equations (\ref{lcdmcont}) and (\ref{lcdmeuler}), which can be manipulated to obtain
 \begin{align}
 \updelta_{\Rm{N}}^{(1)\prime\prime} + \mathcal{H} \updelta_{\Rm{N}}^{(1)\prime} - \frac{3\mathcal{H}_0^2\Omega_{m0}}{2a}\updelta_{\Rm{N}}^{(1)} = 0\;,
 \end{align}
and
\begin{align}\label{uevo}
 U^{(1)\prime\prime} + 3\mathcal{H} U^{(1)\prime} + a^2 \Lambda U^{(1)} = 0 \;.
\end{align}
The first of these two equations is solved by $\updelta_{\Rm{N}}^{(1)} = \mathcal{D}(a) \updelta_{0}^{(1)}(\Bf{x})$, where $\mathcal{D}(a)$ is the standard growth factor and $\updelta_{0}^{(1)}(\Bf{x})$ is the spatial initial condition. The second equation is solved by $U^{(1)} = \varphi = g(\tau) \varphi_0 (\Bf{x})$, where $g(\tau)$ is a time-dependent factor and $\varphi_0 (\Bf{x})$ is the initial condition for the gravitational potential itself. These equations are identical to the ones used in standard Newtonian perturbation theory at linear in a $\Lambda$CDM universe, and we will not dwell on them further here.

We can now find the first approximation to equations (\ref{evollcdm})-(\ref{momentum}), using the linearized solutions to the Newtonian equations given above. This begins by decomposing the cosmological variables into the different orders that will be solved for one-by-one using the solutions from the Newtonian perturbation theory applied to the short-scale domain, and results in the following:
\begin{align} \label{Cpertserieslcdm}
\updelta &= \updelta^{(1)} + \frac{1}{2} \updelta^{(2)} + \dots = \sum_{n=1}^{\infty} \frac{\updelta^{(n)}}{n!} \ ,\\
\theta &= \theta^{(1)} + \frac{1}{2} \theta^{(2)} + \dots  = \sum_{n=1}^{\infty} \frac{\theta^{(n)}}{n!} \ , \\
\psi &= \psi^{(1)} + \frac{1}{2} \psi^{(2)} + \dots = \sum_{n=1}^{\infty} \frac{\psi^{(n)}}{n!}  \ , \\
\phi &= \phi^{(1)} + \frac{1}{2} \phi^{(2)} + \dots =  \sum_{n=1}^{\infty} \frac{\phi^{(n)}}{n!}  \;,
\end{align}
where numbers in brackets again refer to the order of approximation in the Newtonian perturbation theory. These expression can be subsituted into Equations (\ref{evollcdm})-(\ref{momentum}), which immediately result in
\begin{align}
\psi^{(1)} = \phi^{(1)}\;.
\end{align}
The evolution equation for the one degree of freedom that remains in the scalar sector of the gravitational theory can then be found to be
\begin{align}
(\psi^{(1)}+U^{(1)})'' + 3\mathcal{H}(\psi^{(1)}+U^{(1)})' + a^2 \Lambda (\psi^{(1)}+U^{(1)}) = 0 \;.
\end{align}
This equation is identical in form to Equation (\ref{uevo}), and is therefore solved by function with the same time dependency, $g(\tau)$. This fact leads us to consider $\psi^{(1)}+U^{(1)}$ as one single variable, with support on all length scales, which we will henceforth label as $\varphi$. Similarly absorbing the parts of $\theta^{(1)}$ and $\updelta^{(1)}$ that behave like the first-approximations to the corresponding Newtonian quantities into $\theta^{(1)}_{\rm N}$ and $\updelta^{(1)}_{\rm N}$ gives us that only remaining non-Newtonian part of these quantities is given by
\begin{equation}
\updelta^{(1)} = - 2\varphi \; .
\end{equation}
This process of re-defining Newtonian quantities to absorb the long-wavelength parts of the corresponding large-scale perturbations is described in more detail in Equations (5.12)-(5.16) of Ref. \cite{gallagher2020multi}, and proceeds in exactly the same way here.  

These results show that the first approximation to the 2PPT equations gives identical results to standard first-order cosmological perturbation theory in a $\Lambda$CDM universe. Let us now consider the second approximation, which gives more interesting results.

\subsection{Second approximation}

After applying Newtonian perturbation theory to equations (\ref{newtU})-(\ref{lcdmeuler}), we can extract the following evolution equation at second approximation:
\begin{align}
&\;\;\;U^{(2)\prime\prime}  + 3\mathcal{H} U^{(2)\prime} + a^2 \Lambda U^{(2)} 
=  \frac{4}{3}g^2 \bigg( \frac{f^2}{\Omega_m} + \frac{3}{2} \nabla^{-2} \partial_i \partial^j(\partial^i\varphi_0 \partial_j \varphi_0) \bigg) - g^2 (\nabla\varphi_0)^2\;,
\end{align}
where $f=d(\log \mathcal{D})/d(\log a)$ is the growth rate of structure, and where $g$ and $\varphi_0$ are as in the equations presented above. This equation is solved by the usual expression for second-order Newtonian solutions in a $\Lambda$CDM universe \cite{bartolo2006full}:
\begin{align}
U^{(2)} = \frac{2\mathcal{D}^2}{3a\mathcal{H}_0\Omega_{m0}} \partial_i\varphi_0 \partial^i\varphi_0 - \frac{4(\mathcal{D}^2+ \mathcal{F})}{3a\mathcal{H}_0\Omega_{m0}} \Bf{\Psi}_0 \;,
\end{align}
where  we have used the Newtonian kernel
$
\Bf{\Psi}_0 = -\frac{1}{2} \nabla^{-2} \left[ \left( \nabla^2 \varphi_0 \right)^2- \partial_i \partial_j \varphi_0 \partial^i \partial^j \varphi_0 \right]
$,
as well as defining
\begin{align}
\mathcal{F}&= \mathcal{D}^2 \left[ \frac{\Omega_m}{4}- \frac{\Omega_{\Lambda}}{2} - \frac{1}{U_{3/2}} \left[1- \frac{3}{2} \frac{U_{5/2}}{U_{3/2}}\right]\right]
\end{align}
where
\begin{align}
U_{\alpha} &= \int_0^1 dx \left[\frac{\Omega_m}{x}+\Omega_{\Lambda} x+1-\Omega_m-\Omega_{\Lambda} \right]^{-\alpha} \, .
\end{align}

To find the large-scale potentials $\phi$ and $\psi$ at second approximation it is useful to isolate the combination $\psi^{(2)} - \phi^{(2)}$, using the trace-free $ij$ field equation. The cosmological constant does not affect this equation, so the result is identical to the corresponding equation in an Einstein-de Sitter universe \cite{gallagher2020multi}:
\begin{align}
\nabla^2\nabla^2 \psi^{(2)} &= \nabla^2 \nabla^2\phi^{(2)} - 4g^2 \nabla^2\nabla^2 \varphi_0 
-8g^2 \bigg(\frac{f^2}{\Omega_{m}} + \frac{3}{2}  \bigg)\nabla^2\nabla^2 \Bf{\Theta}_0 
-4g^2 \varphi_0^2 + Q\;,
\end{align}
where
\begin{align}
P^i_{\phantom{i} j} &=2 \partial^i \varphi \partial_j \varphi + 8 \pi a^2 \bar{\rho}v^{(1)i} v^{(1)}_{j}\, ,\qquad
\nabla^2 N = \partial_i \partial^j P^i_{\phantom{i} j} \qquad {\rm and} \qquad \nabla^2 Q = -P +3 N \,,
\end{align}
and where $P=P^i_{\phantom{i} i}$. Using this allows us to write the second-approximation to the large-scale equations (\ref{evollcdm})-(\ref{momentum}) as
\begin{align}
\psi^{(2)\prime\prime} + 3\mathcal{H} \psi^{(2)\prime} + a^2 \Lambda \psi^{(2)} 
=&  - 4\mathcal{H}^2g^2 \bigg( 2(f-1) + \Omega_m \Big(1- \frac{1}{\Omega_m} \Big) \bigg) \varphi_0^2 
\nonumber\\ &+ 12 g^2 \mathcal{H}^2 \Omega_m \bigg( 2\frac{(f-1)^2}{\Omega_m} - \frac{3}{\Omega_m} + 3\bigg) \Bf{\Theta}_0 \;, \label{psi2eq}
\end{align}
where we have in addition made use of the results
\begin{align}
-4a^2 \Lambda &= 4 \mathcal{H}^2 \Omega_{m} \bigg(1 - \frac{1}{\Omega_m} \bigg) \;, \qquad
\mathcal{H} Q' + a^2 \Lambda Q = 12g^2 \mathcal{H}^2\Omega_m \bigg(2\frac{(f-1)^2}{\Omega_m} - \frac{3}{\Omega_m} + 3 \bigg) \Bf{\Theta}_0 \;, 
\nonumber \\
N &= \frac{4}{3}g^2 \bigg( \frac{f^2}{\Omega_m} + \frac{3}{2} \nabla^{-2} \partial_i \partial^j(\partial^i\varphi_0 \partial_j \varphi_0) \bigg) \;, \qquad
-8 \mathcal{H} \varphi \varphi' = - 8 \mathcal{H}^2 g^2 (f-1) \varphi_0^2 \;.
\end{align}
If one were to compare Equation (\ref{psi2eq}) to the corresponding equation in standard second-order cosmological perturbation theory \cite{villa2016relativistic}, it would be immediately apparent that there are a number of differences.  In particular the terms $8\mathcal{H} \varphi \varphi' $, $(\varphi')^2$, and  $4a^2 \Lambda \varphi^2$ are all absent here. This difference is due to the two-parameter counting scheme, which at leading-order neglects the effects of terms that are quadratic in large-scale perturbations. Such a difference does not occur in dust-only Einstein-de Sitter cosmologies \cite{gallagher2020multi}, as in that case the potentials are independent of time.

The solutions to Equation (\ref{psi2eq}) can be written as
\begin{align} \label{2pgsol}
\psi^{(2)} = b_1 \frac{\varphi_0^2}{a} + 6 b_2 \frac{\Bf{\Theta}_0}{a} + \frac{g}{g_{in}}\psi^{(2)}_{in} \;, 
\end{align}
where the $b_n$ satisfy
\begin{align}
b^{\prime\prime}_1 + \mathcal{H} b^{\prime}_1 - \frac{3}{2}\frac{\mathcal{H}_0^2\Omega_{m0}}{a}b^{}_1 & = -\frac{4\mathcal{H}^2\mathcal{D}^2}{a}\bigg(2(f-1) + (\Omega_m -1) \bigg) \;, \label{b1} \\
b_2'' + \mathcal{H} b_2' - \frac{3}{2}\frac{\mathcal{H}_0^2\Omega_{m0}}{a}b_2 & = \frac{2\mathcal{H}^2\mathcal{D}^2}{a}\bigg(2(f-1)^2 + 3(\Omega_m -1) \bigg) \;. \label{b2}
\end{align}
The solution for $b_2$ is the same as in second-order cosmological perturbation theory \cite{bartolo2006full}, and is given by
\begin{align}
b_2 = -2 \mathcal{D} (g_{in} - g) \;.
\end{align}
The solution for $b_1$, on the other hand, is given by
\begin{align}
b^{}_1 =  - \mathcal{D} \int_{a_{in}}^a  \frac{\mathcal{I}}{a^2\mathcal{W}} \,\Rm{d}a + \frac{\mathcal{H}}{a}\int_{a_{in}}^{a}  \frac{\mathcal{D} \,\mathcal{I}}{a\mathcal{H}\mathcal{W}} \,\Rm{d}a \;,
\end{align}
where $\mathcal{I}(a)$ and $\mathcal{W}(a)$ are the source and Wronskian functions
\begin{align}
\mathcal{I} = \frac{\mathcal{H}^2\mathcal{D}^2}{a}\bigg( -8f - 4\Omega_m +12\bigg) \qquad {\rm and} \qquad
\mathcal{W} = - \frac{\mathcal{H}^2\mathcal{D}}{a}\bigg( f + \frac{3}{2}\Omega_m \bigg) \;.
\end{align}
This result can be seen to reduce to the Einstein-de Sitter limit as $\Lambda \rightarrow 0$, by noting that $\mathcal{I} \rightarrow 0$ in this case. While this result is cumbersome to deal with analytically, it straightforward to evaluate numerically.

\section{Initital conditions and kernels}
\label{sec:kernels}

Let us now turn to the question of how to set initial conditions at $\tau_{in}$, for the expressions derived in the previous section. For this we choose to proceed by using standard cosmological perturbation theory to model the growth of structure in the early universe (before nonlinear structures developed), and then mapping those perturbations onto our 2PPT variables as some \textit{crossover} time, $\tau_{\Rm{cross}}$. We therefore take
\begin{align}
g_{in} = g(\tau_{\Rm{cross}}) \;.
\end{align}
Now, the function $g(z)$ is known to approach a constant value in matter-domination, which means we are free to select any moment of time as the crossover time, provided that it is well within the matter-dominated era. We choose a redshift of $z=19$ for $z_{\Rm{cross}}$, which is equivalent to $a_{in} = 0.05$. 

\begin{figure}
\centering
\includegraphics[scale=0.67]{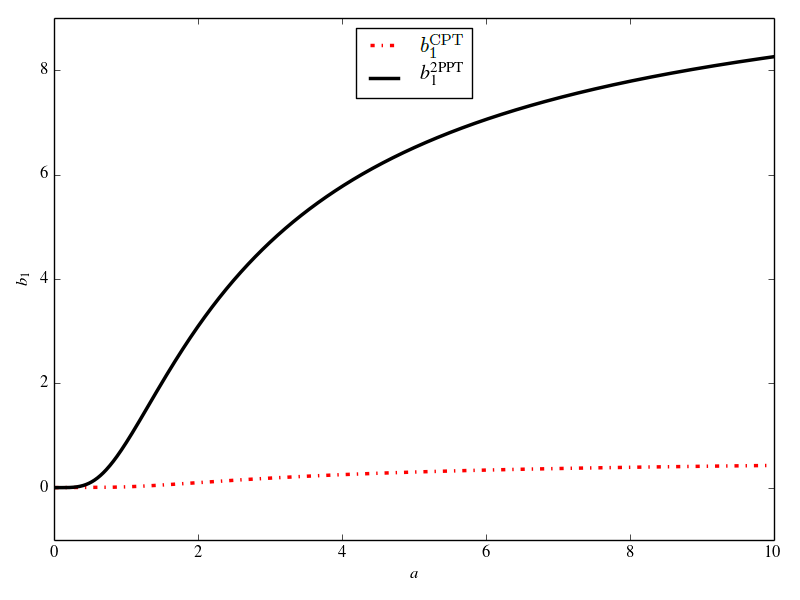}
\vspace{-15pt}
\caption{A plot of $b_1^{\mathrm{2PPT}}=b_1^{\mathrm{2PPT}}(a)$ and $b_1^{\rm CPT}=b_1^{\rm CPT}(a)$, calculated through numerical integration. The two curves can be seen to deviate when $\Lambda$ becomes significant.
}
\label{2pb1}
\end{figure}

In the case of the quantity $b_1$, we can note that we are free to add a homogeneous solution to the differential equation it must obey. We use this fact to ensure continuity of the metric at $\tau_{in}$ by setting 
\begin{align}
b^{\mathrm{2PPT}}_1(a) =  b^{}_1(a)  - \frac{b_1^{}(a_{in}) - b_1^{\rm CPT}(a_{in})}{\mathcal{D}(a_{in})} \mathcal{D}(a) \;,
\end{align}
where $b_1^{\rm CPT}$ is the solution to the equivalent equation for $b_1$ in standard second-order cosmological perturbation theory. This definition ensures that $b^{\mathrm{2PPT}}_1(a_{in}) = b_1^{\rm CPT}(a_{in})$, and therefore that the metric is continuous at the crossover time. We plot $b^{\mathrm{2PPT}}_1$ in Figure \ref{2pb1}, along with $b_1^{\rm CPT}$ for comparison. The two quantities can be seen to differ at late times due to the $\Lambda$-induced time-dependence of the first-order solutions, and the presence of extra terms at second-order in cosmological perturbation theory that contain their time derivatives.

\subsection{Gravitational potential kernel}

In order to calculate the bispectrum of the gravitational potential, we write the solutions from the previous section as
\begin{align}
U^{(2)}+\psi^{(2)}= C_1\varphi_0^2  + C_2 \Bf{\Theta}_0 + C_3 \Bf{\Psi}_0  + C_4 (\nabla \varphi_0)^2 \;,
\end{align}
where
\begin{align}
C_1 &=  2 g g_{in} \bigg(-\frac{5}{3}(a_{\Rm{NL}} -1) -1 \bigg)  + \frac{b_1^{\Rm{2PPT}}}{a}\;, \\
C_2 &= 6\bigg(2g^2 - \frac{10}{3}gg_{in} \bigg)  \;, \\
C_3 &= - \frac{4}{3a \mathcal{H}_0^2 \Omega_{m0}}(\mathcal{D}^2 + \mathcal{F}) \;, \\
C_4 &=  \frac{2}{3a \mathcal{H}_0^2 \Omega_{m0}}\mathcal{D}^2 \;.
\end{align}
These are the same as the corresponding quantities in second-order perturbation theory \cite{bartolo2006full}, apart from $C_1$ which is modified due to the structure of the 2PPT equations. Taking the Fourier transform of this expression we obtain
\begin{align}
U^{(2)}(\Bf{k})+\psi^{(2)}(\Bf{k}) =& \int \Rm{d}^3q_1 \Rm{d}^3q_2 \,\delta^{(3)} (\Bf{k} - \Bf{q}_1 - \Bf{q}_2 )\; \times \nonumber \\
&\quad \Bigg[ C_1 - \frac{C_2}{2k^4}q_1^2 q_2^2 \bigg( 1 - (\Bf{\hat{q}}_1\cdot \Bf{\hat{q}}_2)^2\bigg) - \frac{C_2}{3k^2}\,q_1q_2(\Bf{\hat{q}}_1\cdot \Bf{\hat{q}}_2) \nonumber \\
&\qquad  + \frac{C_3}{2k^2}\,q_1^2 q_2^2 \bigg( 1 - (\Bf{\hat{q}}_1\cdot \Bf{\hat{q}}_2)^2\bigg) + C_4\,q_1q_2(\Bf{\hat{q}}_1\cdot \Bf{\hat{q}}_2)  \Bigg]\varphi_0(q_1) \varphi_0(q_2) \;,
\end{align}
which immediately allows us to write down an expression for the kernel
\begin{align}
\mathcal{K}_2^{U+\psi} =  &\Bigg[ C_1 - \frac{C_2}{2k^4}q_1^2 q_2^2 \bigg( 1 - (\Bf{\hat{q}}_1\cdot \Bf{\hat{q}}_2)^2\bigg) - \frac{C_2}{3k^2}\,q_1q_2(\Bf{\hat{q}}_1\cdot \Bf{\hat{q}}_2) 
\nonumber \\&\qquad  
+ \frac{C_3}{2k^2}\,q_1^2 q_2^2 \bigg( 1 - (\Bf{\hat{q}}_1\cdot \Bf{\hat{q}}_2)^2\bigg) + C_4\,q_1q_2(\Bf{\hat{q}}_1\cdot \Bf{\hat{q}}_2)  \Bigg] \;.
\end{align}
Let us know consider the kernel for matter perturbations.

\subsection{Dark matter kernel}

\begin{figure}
\centering
\includegraphics[scale=0.67]{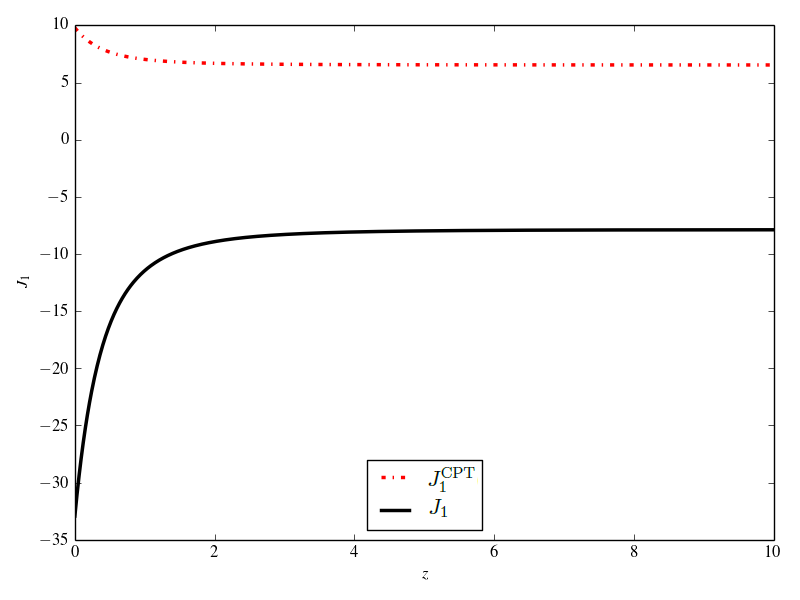}
\vspace{-15pt}
\caption{A plot of $J_1(z)$, and the corresponding quantity $J^{\rm CPT}_1(z)$ from standard second-order cosmological perturbation theory.
}
\label{2pJ1}
\end{figure}

\begin{figure}
\centering
\includegraphics[scale=0.67]{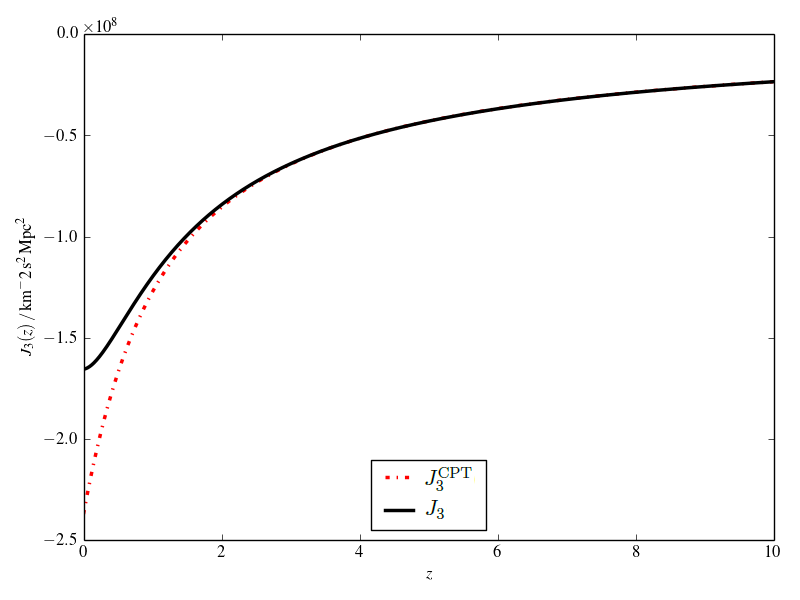}
\vspace{-15pt}
\caption{A plot of $J_3(z)$, and the corresponding quantity $J^{\rm CPT}_3(z)$ from second-order cosmological perturbation theory. 
}
\label{2pJ3}
\end{figure}

\begin{figure}
\centering
\includegraphics[scale=0.67]{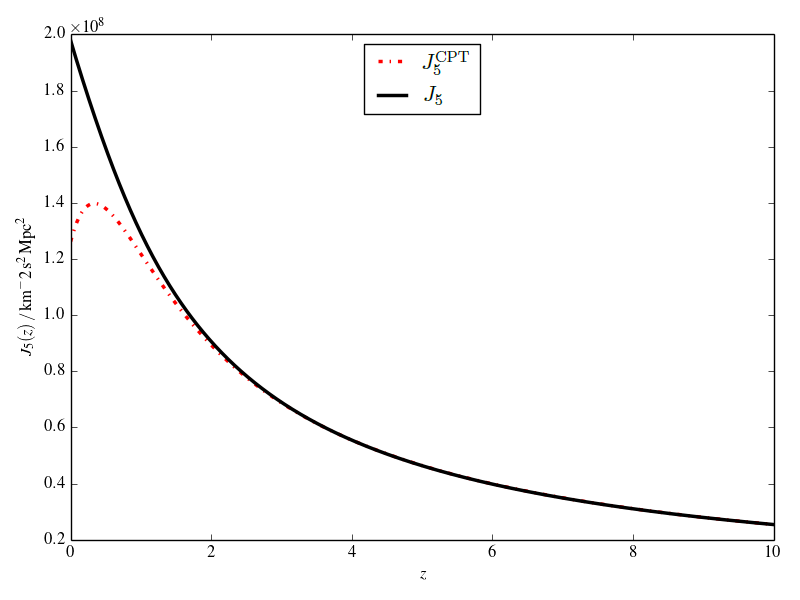}
\vspace{-15pt}
\caption{A plot of $J_5(z)$ solution, and the corresponding quantity $J^{\rm CPT}_5(z)$ from second-order cosmological perturbation theory. 
}
\label{2pJ5}
\end{figure}

Given all the constituent parts of our solution in Equation (\ref{2pgsol}), we can proceed by using the following expression to determine $\updelta^{(2)}$:
\begin{align}
\frac{1}{3}\nabla^2 \psi^{(2)} -&\mathcal{H}(\psi^{(2)} + U^{(2)})' -  \mathcal{H}^2(\psi^{(2)} + U^{(2)}) \nonumber \\
& = \frac{1}{2} \frac{\mathcal{H}^2_0\Omega_{m0}}{a}\updelta^{(2)} + \frac{\mathcal{H}^2_0\Omega_{m0}}{a}(v_{N}^{(1)})^2 - \frac{8}{3}g^2\varphi_0 \nabla^2 \varphi_0 - g^2 \nabla (\varphi_0)^2 \;.
\end{align} 
After some manipulation, we find that this equation implies
\begin{align} \label{d2eq}
\delta^{(2)}_{\rm N}+\delta^{(2)} = J_1(\tau) \varphi_0^2 + J_2(\tau) \Bf{\Theta}_0 + J_3(\tau)(\nabla\varphi_0)^2 + J_4(\tau)\mathcal{F} + J_5(\tau) \varphi_0\nabla^2\varphi_0 \;,
\end{align}
where 
\begin{align} \label{j1eq}
J_1&=  \Bigg[ -\frac{4\mathcal{H} \mathcal{D}'}{\mathcal{H}_0^2\Omega_{m0}} g_{in} \bigg(-\frac{5}{3}(a_{\Rm{NL}}-1) -1 \bigg) 
-\frac{8\mathcal{H}^2\mathcal{D}^2}{a\mathcal{H}_0^2 \Omega_{m0}} - \frac{2\mathcal{H}}{\mathcal{H}_0^2\Omega_{m0}} (b^{\mathrm{2PPT}}_1)' \Bigg] \;, \\
J_2 &=  -\frac{24 {\mathcal H}  {\mathcal D}' {\mathcal D}}{a {\mathcal H}_0^2 \Omega_{\rm m0}}  \;, \\
J_3 &= \Bigg[-\frac{4}{9} \frac{\mathcal{D} g_{in}}{\mathcal{H}_0^2\Omega_{m0}} (1 + 10 a_{\Rm{NL}} ) + \frac{10}{3} \frac{\mathcal{D} g}{\mathcal{H}_0^2 \Omega_{m0}} 
-\frac{8 (\mathcal{D}')^2 }{9\mathcal{H}_0^4 \Omega_{m0}^2} + \frac{4 }{\mathcal{H}_0^2 \Omega_{m0}}b^{\mathrm{2PPT}}_1 \Bigg]\;,\\
  J_4 &=  \;\;\frac{8 {\mathcal H}  {\mathcal F}'}{3({\mathcal H}_0^2 \Omega_{\rm m0})^2} \;, \\
J_5 &= \Bigg[ \frac{8 \mathcal{D} g_{in}}{3 \mathcal{H}_0^2\Omega_{m0}}  \bigg(-\frac{5}{3}(a_{\Rm{NL}}-1) -1 \bigg) 
+ \frac{16}{3} \frac{\mathcal{D} g}{\mathcal{H}_0^2 \Omega_{m0}} + \frac{4 }{\mathcal{H}_0^2 \Omega_{m0}}b^{\mathrm{2PPT}}_1 \Bigg] \;. \label{j5eq}
\end{align}
The functions $J_1$,  $J_3$ and  $J_5$ differ from the corresponding quantities determined using second-order cosmological perturbation theory, and are plotted in Figures \ref{2pJ1}, \ref{2pJ3} and \ref{2pJ5} for comparison. It may be noted that the expected Einstein-de Sitter behaviour is recovered at high redshift, up to a mathematical normalization that is physically unimportant.

The solution from Equation (\ref{d2eq}) can be expressed as
\begin{align}
\delta^{(2)}_{\rm N}(k,\tau)+\delta^{(2)}(k,\tau) = & \; \int \Rm{d}^3 q_1\, \Rm{d}^3 q_2\, \delta^{(3)}(\Bf{k} - \Bf{q}_1 -\Bf{q}_2)\,\delta^{(1)}_{\rm N}(\Bf{q}_1)\delta^{(1)}_{\rm N}(\Bf{q}_2) \mathcal{K}_{2}(\Bf{q}_1, \Bf{q}_2,\tau)\;,
\end{align}
where the kernel $\mathcal{K}_{2}(\Bf{k}_1, \Bf{k}_2)$ is given by
\begin{align}
 \mathcal{K}_{2}(\Bf{k}_1, \Bf{k}_2) = \frac{\big(\beta - \alpha \big) + \frac{\beta}{2} \,\hat{\Bf{k}}_1 \cdot \hat{\Bf{k}}_2 \Big( \frac{k_1}{k_2} + \frac{k_2}{k_1} \Big) + \alpha \Big(\hat{\Bf{k}}_1 \cdot \hat{\Bf{k}}_2\Big)^2 + \gamma \Big( \frac{k_1}{k_2} - \frac{k_2}{k_1} \Big)^2}{ \bigg( 1 + \frac{3\mathcal{H}^2f}{q_1^2}\bigg) \bigg( 1 + \frac{3\mathcal{H}^2f}{q_2^2} \bigg) }\;,
 \end{align}
 and where
  \begin{align}
 \alpha(k,\tau) &=  \bigg(1- \frac{\mathcal{F}}{\mathcal{D}^2}\bigg) +\frac{ (\mathcal{H}_0^2\Omega_{m0})^2}{\mathcal{D}^2}\Bigg[-\frac{9(4J_3+J_4)}{8k^2}  + \frac{9  J_1}{k^4} - \frac{3 J_2}{8k^4}\Bigg]\;, \\
 \beta(k,\tau) &=  2 + \frac{ (\mathcal{H}_0^2\Omega_{m0})^2}{\mathcal{D}^2}\Bigg[- \frac{9(J_3+J_5)}{2k^2} + \frac{18 J_1}{k^4} - \frac{3J_2}{2k^4} \Bigg] \;,  \\
 \gamma(k,\tau) &= \frac{ (\mathcal{H}_0^2\Omega_{m0})^2}{\mathcal{D}^2}\Bigg[- \frac{9J_5}{8k^2} + \frac{9J_1}{4k^4}\Bigg] \;.
 \end{align}
These expressions are identical to those from second-order cosmological perturbation theory \cite{tram2016intrinsic}, except that in this case the functions $J_n$ are given by Equations (\ref{j1eq})-(\ref{j5eq}). These expressions enable the calculation of the correlation functions of the dark matter overdensity.

\section{Calculation of the matter bispectrum}
\label{bispectrumsec}

In this section, we will present the results of the leading order corrections to the bispectra of the metric potential $\psi^{(2)}+ U^{(2)}$ and the dark-matter overdensity $\updelta^{(2)} + \updelta_{\Rm{N}}^{(2)}$.

\subsection{Gravitational potentials}

\begin{figure}
\centering
\includegraphics[scale=0.67]{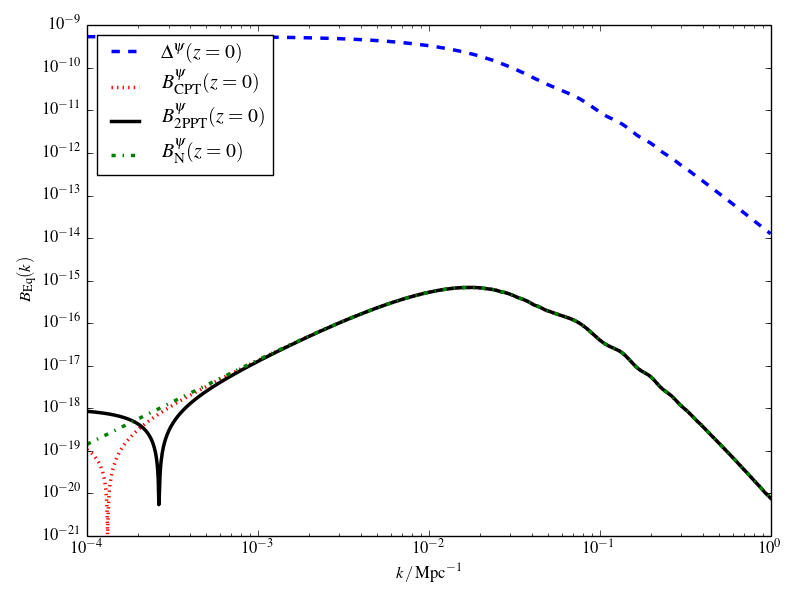}
\vspace{-15pt}
\caption{A plot of the equilateral configuration of the bispectrum of gravitational potentials in 2PPT at redshift $z=0$, as compared to Newtonian theory and second-order CPT.}
\label{PsiEqz0}
\end{figure}

\begin{figure}
\centering
\includegraphics[scale=0.67]{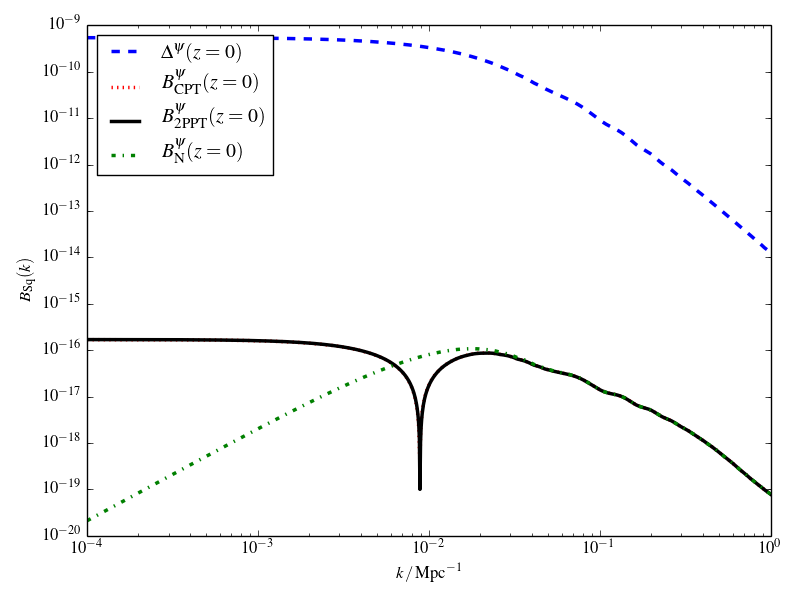}
\vspace{-15pt}
\caption{A plot of the squeezed configuration of the bispectrum of gravitational potentials in 2PPT at redshift $z=0$, as compared to Newtonian theory and second-order CPT.
}
\label{PsiSqz0}
\end{figure}

\begin{figure}
\centering
\includegraphics[scale=0.67]{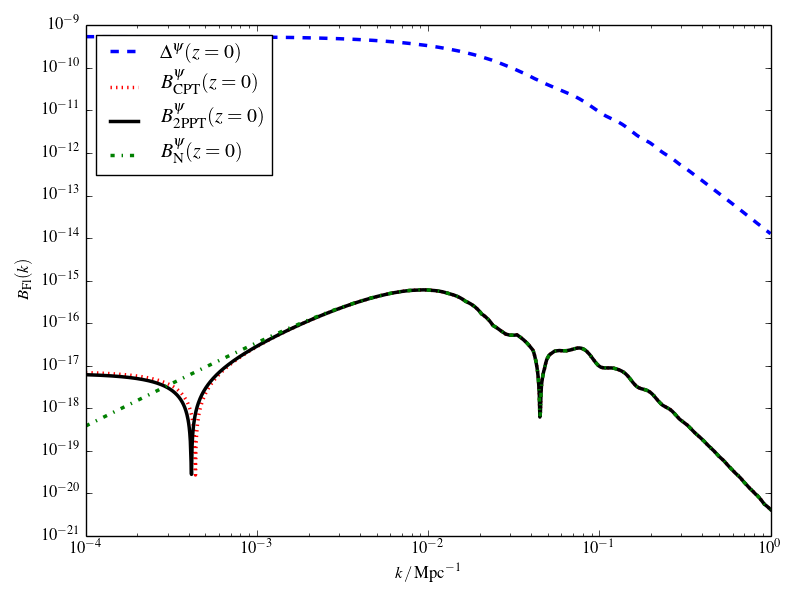}
\vspace{-15pt}
\caption{A plot of the flatenned configuration of the bispectrum of gravitational potentials in 2PPT at redshift $z=0$, as compared to Newtonian theory and second-order CPT.
}
\label{PsiFlz0}
\end{figure}

The leading order contribution to the dimensionless bispectrum of the scalar metric potential $\psi+U$ is calculated as follows:
\begin{align}
&\langle (\psi+ U) (\psi+ U) (\psi+ U) \rangle \nonumber \\ \simeq  &\langle (\psi^{(2)}+ U^{(2)}) (\psi^{(1)}+ U^{(1)}) (\psi^{(1)}+ U^{(1)}) \rangle \nonumber \; \\
 = &(2\pi)^{3} \delta^{(3)}(\Bf{k}-\Bf{q}_1 -\Bf{q}_2) \mathcal{K}^{\psi + U}_2(\Bf{q}_1,\Bf{q}_2,\Bf{k}) P^{\psi}(q_1)P^{\psi}(q_2) + \;2\;\Rm{cycl.}\;\Rm{perms} \;, 
\end{align}
where $P^{\psi}(k)$ is the dimensional linear power spectrum of the gravitational potential, as predicted by \textit{CLASS} \cite{lesgourgues2011cosmic1, blas2011cosmic2, lesgourgues2011cosmic3, lesgourgues2011cosmic4}. The implication of this result is that we can write the dimensionless bispectrum $B^{\psi+U}$ in terms of the kernel $\mathcal{K}^{\psi + U}_2$ as 
\begin{align}
B^{\psi+U}(k_1,k_2,k_3) =  \mathcal{K}^{\psi + U}_2(\Bf{q}_1,\Bf{q}_2,\Bf{k}) \Delta^{\psi}(q_1)\Delta^{\psi}(q_2) + \;2\;\Rm{cycl.}\;\Rm{perms} \;,
\end{align}
where $\Delta^{\psi} = \frac{2\pi^2}{k^3}P^{\psi}(k)$. We plot the dimensionless bispectrum of gravitational potentials at redshift $z=0$ in various different types of triangular configurations in Figures \ref{PsiEqz0}, \ref{PsiSqz0} and \ref{PsiFlz0}. These results are plotted together with the linear power spectrum $\Delta^{\psi}$ in order to give the reader a reference point for the magnitudes of the quantities involved. 


These plots demonstrate how the 2PPT solution for the gravitational potential deviates from the Newtonian prediction at small values of $k$. This behaviour can be anticipated from the form of the integral kernel; Newtonian terms come with an extra $k^2$ and hence dominate at large $k$, whereas relativistic corrections are dimensionless, and hence are only visible at small values of $k$. The difference with cosmological perturbation theory solution is maximised in the equilateral configuration, whilst it is significantly reduced in the flattened case.

In the case of the squeezed limit shown in Figure \ref{PsiSqz0}, it is clear that the differences between the 2PPT solution and the second-order CPT solution is negligible even at scales of $k \sim 10^{-4}$. This is because the kernel is dominated by the terms proportional to $\Bf{\hat{q}}_1 \cdot \Bf{\hat{q}}_2$ in this limit, and these terms are the same in both the cosmological perturbation theory and 2PPT case. We have verified that differences between 2PPT and CPT become negligible at higher redshifts, even in the equilateral configuration, as in this case we move into the matter-dominated era where the differences between the gravitational potentials in each approach become negligible.

\subsection{Dark matter overdensity}

\begin{figure}
\centering
\includegraphics[scale=0.67]{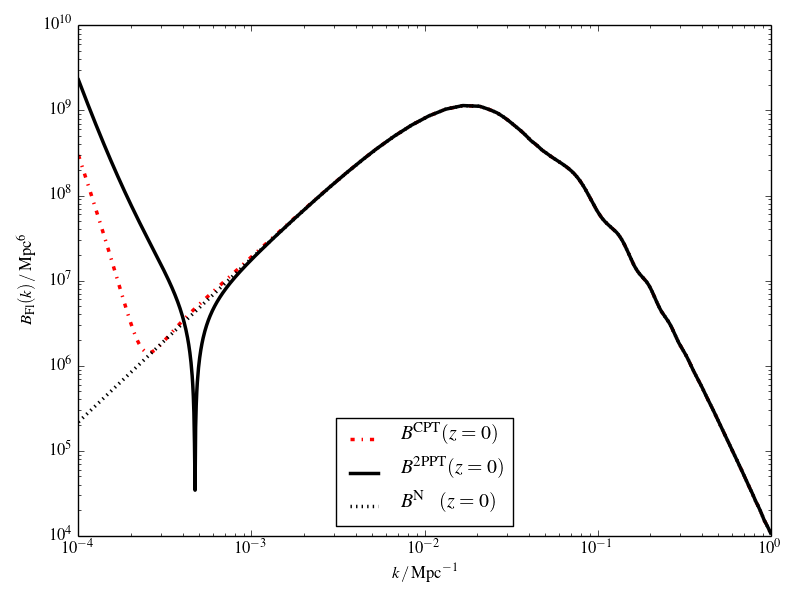}
\vspace{-15pt}
\caption{A plot of the equilateral configuration of the bispectrum of dark matter in 2PPT at redshift $z=0$, as compared to Newtonian theory and second-order CPT.}
\label{Eq}
\end{figure}

\begin{figure}
\centering
\includegraphics[scale=0.67]{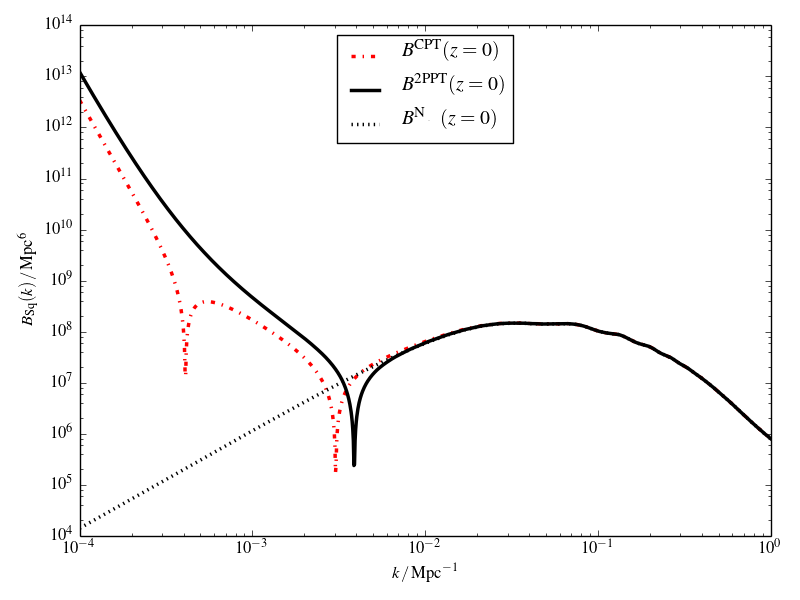}
\vspace{-15pt}
\caption{A plot of the squeezed configuration of the bispectrum of dark matter in 2PPT at redshift $z=0$, as compared to Newtonian theory and second-order CPT. 
}
\label{Sq}
\end{figure}

\begin{figure}
\centering
\includegraphics[scale=0.67]{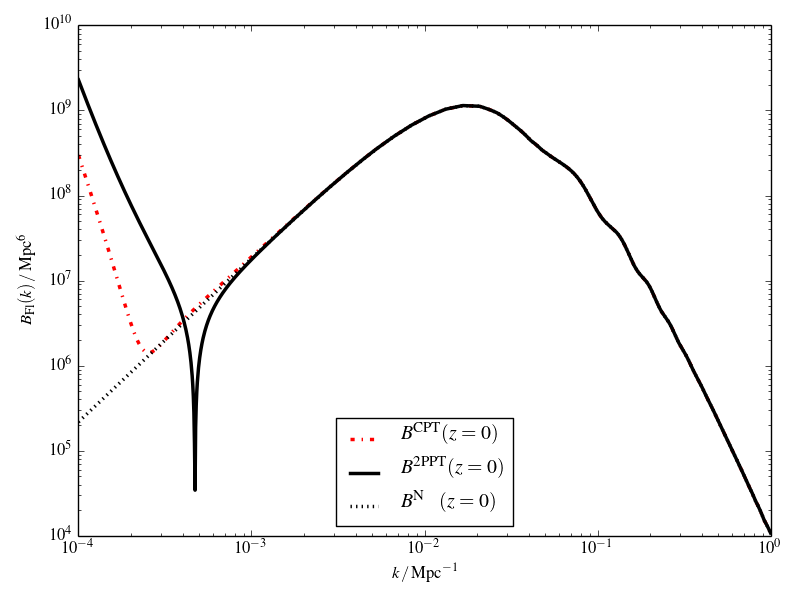}
\vspace{-15pt}
\caption{A plot of the flattened configuration of the bispectrum of dark matter in 2PPT at redshift $z=0$, as compared to Newtonian theory and second-order CPT.
}
\label{Fl}
\end{figure}

The second approximation to the dark-matter overdensity leads to the following expression for the leading order bispectrum in our approach:
\begin{align}
B^{\Rm{2PPT}}(k_1,k_2,k_3) =  \mathcal{K}^{\Rm{2PPT}}_{2}(\Bf{k}_1, \Bf{k}_2)P(k_1)P(k_2) + \; \Rm{cycl} \; \Rm{perms}\;.
\end{align}
This quantity is plotted at redshift $z=0$ in Figures \ref{Eq}, \ref{Sq} and \ref{Fl}, in the equilateral, squeezed and flattened configurations respectively.

Figures \ref{Eq} \ref{Sq} and \ref{Fl} display more significant deviation from the results of second-order CPT than in the Einstein-de Sitter case examined in Ref. \cite{gallagher2020multi}. This is to be expected, as the differences in the field equations are more significant. In the Einstein-de Sitter case, time derivatives of the first-order solution vanished, meaning that the omission of time-dependent terms like $\varphi'^2$ and $a^2 \mathcal{H}^2\varphi^2$ were not a factor. However, these terms are non-zero in $\Lambda$CDM. The fact that they are present in the second-order CPT equations, but not the 2PPT field equations, therefore leads to the deviations shown in the plots.

\section{Discussion}
\label{sec:dis}

We have studied the application of the 2-parameter perturbation theory (2PPT) approach to a $\Lambda$CDM universe. This has involved solving for the short-scale Newtonian gravitational potentials using the leading-order part of the 2PPT-expanded field equations, and then using these solutions to determine the effects of nonlinear structure on long-wavelength cosmological fluctuations. We have, in particular, calculated the dark matter bispectrum, and have compared the results to standard second-order cosmological perturbation theory.

The method we used to solve the small-scale equations is Newtonian perturbation theory. This method allows analytically tractable solutions to be found in the quasi-nonlinear regime, where the equations of Newtonian physics on an expanding background are expected to accurately describe the physics, but where perturbation theory is still expected to give a realiable approximation to the true behaviour of the system. This approach has the very considerable benefit of allowing solutions to be written down in a relatively straightforward fashion, but suffers from the drawback that it does not accurately model the gravitational fields of matter on scales where it is highly nonlinear (i.e. on scales $\lesssim 10 \, {\rm Mpc}$).

With solutions for the Newtonian part of the gravitational field in hand, we then solved for the long-wavelength cosmological perturbations. We found that the leading-order contributions to the field equation that governs the evolution of these perturbations, (\ref{evollcdm}), is similar in form to the corresponding equation in second-order cosmological perturbation theory, though not identical. This is in contrast to the case where $\Lambda=0$, where no such difference exists in the evolution equation \cite{gallagher2020multi}. The 2PPT equations are missing terms that would have occurred if calculating the corresponding quantities in second-order cosmological perturbation theory, and the bispectra that result are consequently modified (as shown in Figs. \ref{PsiEqz0} - \ref{Fl}).

Our findings illustrate both the advantages and disadvantages of using the 2PPT approach, as compared with other approaches to modelling structure in the Universe. First of all, our results show that the effects of the quasi-nonlinear Newtonian regime on cosmological fluctuations do indeed faithfully reproduce some of the known effects that are included in second-order cosmological perturbation theory. However, our study also shows that there exist large-scale relativistic effects that are not accounted for in the leading-order 2PPT equations that we have studied here, and that the absense of the terms that correspond to these effects can have considerable consequences for the calculation of observables such as the matter bispectrum on large scales.

The occurence of deviations from second-order cosmological perturbation theory is expected. This is because second-order perturbation theory is a theory of second-order corrections to the Einstein field equations in a regime where all fluctuations are small with respect to the background. The 2PPT field equations instead describe the evolution of first-order perturbations on top of a background which itself contains nonlinear Newtonian structure on small scales. As such, 2PPT will necessarily not contain the same gravitational self-interactions as a second-order theory. Such terms could be included by going to higher order within the two-parameter expansion, or by selectively including terms which can be shown to have significant consequences for a given observable. Our calculation of the bispectrum shows that this may be necessary for at least some observables.

Instead, the strength of the 2PPT approach is that it can include the consequences of nonlinear much more readily. One way of doing this, in the context of the present study, would be to go to higer-order approximations in the Newtonian perturbation theory. This would be significantly easier than going to higher than second-order in cosmological perturbation theory, and could well provide a way of probing deeper into the regime of nonlinear structure formation. A more direct route, though less analytically tractible, would be to use the numerical output of Newtonian cosmological $n$-body simulations as the solutions to the Newtonian sector of the 2PPT system. This would allow for the full nonlinear regime to be included in ways that would be simply impossible in cosmological perturbation theory, and could lead to significant departures from the predictions of second-order perturbation theory on all scales, not just in the non-perturbative regime. We leave the study of these ideas for future work.

\section*{Acknowledgements}

We acknowledge financial support from the STFC under grant ST/P000592/1.

\newpage


\end{document}